\documentclass{article}
\usepackage{graphicx}
\usepackage{epstopdf, epsfig}
\usepackage[sorting=none]{biblatex}
\usepackage[margin=1.5in]{geometry}

\title{An atomistic model for the thermal resistance of a liquid-solid interface}

\author{N. G. Hadjiconstantinou and M. M. Swisher\\
Department of Mechanical Engineering\\
Massachusetts Institute of Technology\\
Cambridge, MA 02139, USA}

\begin{document}

\maketitle

\begin{abstract}
The thermal resistance associated with the interface between a solid and a liquid is analyzed from an atomistic point of view. Partial evaluation of the associated Green-Kubo integral elucidates the various factors governing heat transport across the interface and leads to a quantitative model for the thermal resistance in terms of atomistic-level system  parameters. The model is validated using molecular dynamics simulations.
\end{abstract}

\section{Introduction}
\label{intro}
The thermal resistance associated with the interface between two materials is of great fundamental as well as practical importance \cite{Cahill,Beskok}. In its most general form, it is defined by the relation 
\begin{equation}
    R_K=\frac{\Delta T}{q}
    \label{R_K_def}
\end{equation}
where $q$ denotes the heat flux through the interface and $\Delta T$ the temperature jump across the interface. A large resistance is beneficial for insulating purposes, while a small resistance is desirable in applications involving cooling and heat dissipation in general. 

In this article we focus on the interface between a solid and a liquid at conditions of typical engineering interest (no superfluidity effects). As shown before \cite{chicou,Beskok,Keblinski2011,kimbo,Blady} and explicitly predicted by the model developed here, this resistance is inversely related to the liquid-solid interaction strength and can become quite  appreciable when this interaction is weak. As in the case of its counterpart interfacial transport phenonenon, namely fluid slip at a liquid-solid interface \cite{chicou,Jerryslip,BocquetAnnu,jfm2021}, the thermal resistance becomes even more important at the nanoscale, where surface effects dominate \cite{chicou,Cahill,BocquetAnnu}.  

Heat transfer across a liquid-solid interface is rather poorly understood compared to  solid-solid interfaces or superfluid helium in contact with a solid, both of which have received significant attention  \cite{pohl}. Of note is the case of a dilute gas (in contact with a solid) where a mesoscopic description for transport, namely the Boltzmann equation \cite{NGHpof,Sone}, exists; this description is sufficiently tractable to allow the use of elegant, in the authors' opinion, boundary layer analyses \cite{Sone} to rigorously characterize the inhomogeneity introduced by the solid and the resulting thermal resistance associated with the interface. Such analyses can also be applied to solid-solid interfaces when conditions allow the treatment of the underlying phonon transport by a Boltzmann transport equation \cite{Peraud}. As expected, the phonon picture  plays a prominent role in models of the thermal resistance for solid-solid interfaces \cite{Cahill,ase} even beyond Boltzmann transport. Two of the most well known {\it approximate} models for this resistance, namely the diffuse mismatch and acoustic mismatch model, are based on this picture.

Unfortunately, the above developments have only contributed to a qualitative understanding of transport at the liquid-solid interface, since neither  dilute-gas approximations, nor phonon-based arguments lend themselves naturally to the physical picture of transport in a dense liquid. 
In the present paper we develop a model for the thermal resistance for a liquid-solid interface by partial evaluation of the Green-Kubo (GK) relation in Ref. \cite{chicou}. The resulting model highlights the various factors governing heat transport across the interface and leads to a {\it quantitative description} of the thermal resistance in terms of atomistic-level system  parameters.

Following common practice, in what follows we will refer to $R_K$ as the Kapitza resistance, even though we note that this term has its origins in studies involving superfluid helium. With this in mind, we would like to restate that our interest and analysis here centers on conditions of typical practical interest and not superfluid conditions which have been studied extensively in the physics literature \cite{pohl}.  

The proposed model is developed and discussed in section \ref{model} and validated using MD simulations in section \ref{MD}. We conclude with a discussion of our findings and possible future directions for research in section \ref{discussion}.  

\section{Theoretical Model}
\label{model}
\subsection{Background}
We consider an atomic liquid in contact with an atomic solid. The Kapitza resistance associated with the interface between the two materials is given \cite{chicou,Alosious} by the following GK relation 
\begin{equation}
R_K=\frac{A k_B T^2}{\int_0^\infty \langle q(t)q(0)\rangle dt}.
\label{R_K_equation}
\end{equation}
 In this expression, angled brackets denote ensemble average, $A$ denotes the interface area, $k_B$ is Boltzmann's constant and $T$ is the (interface) temperature. Moreover,
\begin{equation}
q(t)=\sum_{i\in fluid} \sum_{j \in solid} {\bf v}_i \cdot{\bf F}_{ij}
\label{q-def}
\end{equation}
is the energy flux across the interface, where ${\bf v}_i$ is the velocity of fluid particle $i$ and ${\bf F}_{ij}$ is the force exerted by wall atom $j$ on fluid atom $i$. Denoting the total force exerted by the wall on fluid particle $i$ by ${\bf F}_{wi}$, we can write (\ref{q-def}) as 
   \begin{equation}
q(t)=\sum_i {\bf v}_i \cdot{\bf F}_{wi}.
\label{v2}
\end{equation}
\subsection{Model development}
One of the advantages of GK formulations is that they connect transport (non-equilibrium) to the decay of correlations in equilibrium. We exploit this property to  obtain a model for the Kapitza resistance by  using {\it equilibrium} statistical mechanics to partially evaluate (\ref{R_K_equation}) and express it in terms of more accessible system properties.  

We start by writing the heat flux autocorrelation integral in  (\ref{R_K_equation}) in the form
\begin{equation}
\int_0^\infty \langle q(t)q(0)\rangle dt=\langle q(0)q(0)\rangle\int_0^\infty \psi(t)dt,
\end{equation}
where 
\begin{equation}
\psi(t)=\frac{\langle q(t)q(0)\rangle}{\langle q(0)q(0)\rangle}
\label{psi-def}
\end{equation}
and from expression (\ref{v2})
\begin{equation}
\langle q(0)q(0)\rangle=\langle \sum_i {\bf v}_i \cdot{\bf F}_{wi} \sum_j {\bf v}_j \cdot{\bf F}_{wj} \rangle.
\end{equation}
Using the fact that the velocities of different particles as well as different components of a particle's velocity are uncorrelated in equilibrium we obtain
\begin{eqnarray}
\langle q(0)q(0)\rangle&=&\langle \sum_i ({v^2_x}_i {F^2_{wx}}_i+{v_y^2}_i {F^2_{wy}}_i +{v^2_z}_i {F^2_{wz}}_i) \rangle\\&=&\frac{k_B T}{m} \langle \sum_i ({F^2_{wx}}_i+{F^2_{wy}}_i +{F^2_{wz}}_i) \rangle=\frac{k_BT}{m}\langle \sum_i|{\bf F}_w|^2_i\rangle,
\end{eqnarray}
where in the above the subscripts $x,y,z$ denote the three cartesian components and $|{\bf F}_w|_i$ is the magnitude of the force exerted by the wall on particle $i$. The final expression was obtained by noting that, in equilibrium, velocities are uncorrelated to forces and 
\begin{equation}
\langle v_x^2\rangle=\langle v_y^2\rangle=\langle v_z^2\rangle=\frac{k_BT}{m},
\end{equation}
where $m$ denotes the liquid atom mass.
Using the definition 
\begin{equation}
I=\lim_{t\rightarrow \infty}I(t)=\lim_{t\rightarrow \infty}\int_0^t \psi(t')dt'
\label{I-def}
\end{equation}
we obtain the following expression for the Kapitza resistance
\begin{equation}
R_K=\frac{ m T }{ A^{-1}\langle \sum_i|{\bf F}_w|^2_i\rangle I}.
\label{inter}
\end{equation}
This expression shows that, in addition to the explicit dependence on temperature and atomic mass, the Kapitza resistance is primarily determined by two main factors:  the (ensemble averaged) total square wall force (TSWF) per unit area associated with the fluid-solid interaction, $A^{-1} \langle \sum_i|{\bf F}_w|^2_i\rangle$,  and the correlation decay time as measured by the integral $I$ defined in (\ref{I-def}). 

This decomposition of (\ref{R_K_equation}) is important because it allows further progress by using the fact that the two main factors just identified can be treated separately: the TSWF is clearly strongly dependent on the fluid-solid interaction, while $I$ may be expected to be an intrinsic function of the fluid state and properties and thus not be sensitive to the fluid-solid interaction. 
Our MD simulations, detailed in section \ref{MD}, also support this view. In fact, we found $I$ to be quite insensitive overall, which allows us to express it, with small error, in terms of the fluid's atomistic-level thermal relaxation timescale, 
\begin{equation}
    \tau=\frac{\kappa}{M_\infty},
    \label{tau}
\end{equation}
which is also quite insensitive to the fluid state, at least for the simple fluids considered here. In (\ref{tau}), $\kappa$ denotes the fluid thermal conductivity, 
\begin{equation}
    M_\infty=\frac{V}{3 k_B T^2}\langle {\bf J}_q^2\rangle
\end{equation}
denotes the fluid thermal modulus \cite{Heyes2019} and ${\bf J}_q$ denotes the volume-averaged heat-flux vector in a homogeneous fluid  of volume $V$ \cite{Heyes}. The connection between $\tau$  and $I$ can be seen from the GK relation for the thermal conductivity of a homogeneous fluid
\begin{equation}
    \kappa = \frac{V}{3k_BT^2} \int _0^\infty \langle {\bf J}_q(t)\cdot {\bf J}_q(0)\rangle \;dt= M_\infty \int _0^\infty \frac{\langle {\bf J}_q(t)\cdot {\bf J}_q(0)\rangle}{\langle {\bf J}_q(0)\cdot{\bf J}_q(0)\rangle} \;dt.
\end{equation}
Here, with the fluid thermal conductivity known, this relation is used to provide a measure of the relaxation time $I$, via the relation $\tau=D I$, where $D$ is a proportionality constant.

The above allows us to write (\ref{inter}) into the following form, which will serve as our model for the Kapitza resistance
\begin{equation}
R_K=\frac{D m T M_\infty}{\kappa A^{-1} \langle  \sum_i|{\bf F}_w|^2_i\rangle }.
\label{final}
\end{equation}
The value of the coefficient $D$ will be determined with the help of MD simulations in section \ref{MD}, which will also be used to validate the approximation $I= \kappa/D M_\infty$ as well as the full expression (\ref{final}). 
\subsection{The total square wall force}
\label{TSWF}
The mean square force on an atom in a homogeneous fluid has been studied in the past \cite{Hansen} and been shown to be related to the Einstein frequency of the material, $\Omega_0$, via the relation $$\langle |{\bf F}|^2\rangle=3mk_B T\Omega_0^2.  $$
In a homogeneous fluid, $\Omega_0^2$ is linked to the oscillation frequency of the fluid atoms at equilibrium and is a measure of the restoring force associated with their interaction. 

Unfortunately, knowledge of $\langle|{\bf F}|^2\rangle$ is not useful here; the boundary presence introduces an inhomogeneity and makes $|{\bf F}_w|_i$ dependent on the position of particle $i$.  An approximation for the force on particle $i$ located a distance $z$ from the interface can be obtained by integrating over the contributions of all solid atoms (half space defined in cylindrical polar coordinates by $0\leq z'< \infty$, $0\leq R< \infty$), 
\begin{equation}
|{\bf F}_w|(z)=\int_0^\infty \int_0^\infty n_w \left[\frac{du_{wf}(r)}{dr}\right]^2 2\pi R dR dz',\;\;\;\;r=\sqrt{(z+z')^2+R^2},
\label{approx_fsq}
\end{equation} 
where $n_w$ denotes the wall number density.
For the generalized Lennard-Jones potential \cite{Alosious} studied here 
\begin{equation}
    u_{ij}(r)=4\varepsilon_{ij}\left[\left(\frac{\sigma_{ij}}{r}\right)^{12}-C_{ij}\left(\frac{\sigma_{ij}}{r}\right)^6\right],
\end{equation}
where $r$ denotes the distance between atoms $i$ and $j$, the above expression yields 
\begin{equation}
|{\bf F}_w|(z)=2304\pi \varepsilon_{wf}^2 n_w\left[\frac{\sigma_{wf}^{24}}{12\times 23 z^{23}}-\frac{C_{wf}\sigma_{wf}^{18}}{9\times 17 z^{17}}+\frac{C_{wf}^2\sigma_{wf}^{12}}{24\times 11 z^{11}}\right],
\end{equation}
where the subscripts $f$ and $w$ denote the fluid and wall, respectively.
To calculate the overall contribution of fluid atoms, we integrate over the fluid region ($z\ge 0$) to obtain 
\begin{equation}
\langle \sum_i|{\bf F}_w|^2_i\rangle\approx2304\pi \varepsilon_{wf}^2 n_w A \int_0^\infty n(z) \left[\frac{\sigma_{wf}^{24}}{12\times 23 z^{23}}-\frac{C_{wf}\sigma_{wf}^{18}}{9\times 17 z^{17}}+\frac{C_{wf}^2\sigma_{wf}^{12}}{24\times 11 z^{11}}\right] dz,
\label{force-approx}
\end{equation}
where $n(z)$ denotes the (number) density distribution in the fluid region, which is unfortunately unknown. Due to the rapidly decaying nature of the (square of) the interatomic force, non-negligible contributions to this integral come only from the fluid close to the interface, usually referred to as the first fluid layer. The density distribution in this layer is expected to be a function of the bulk fluid density and temperature, as well as other system parameters \cite{Jerryinterface} and needs to be known, or approximated before this relation can be evaluated. Moreover, for this approximate scaling relation to take into account the effect of solid wall structure, the spatial distribution of wall particles needs to be taken into account in (\ref{approx_fsq}). On the other hand, we expect expression (\ref{force-approx}) to be a useful guide on the dependence of the TSWF on system parameters and its concomitant effect on the Kapitza resistance. This will be verified in the following section, where MD simulations will be providing accurate results for the TSWF.

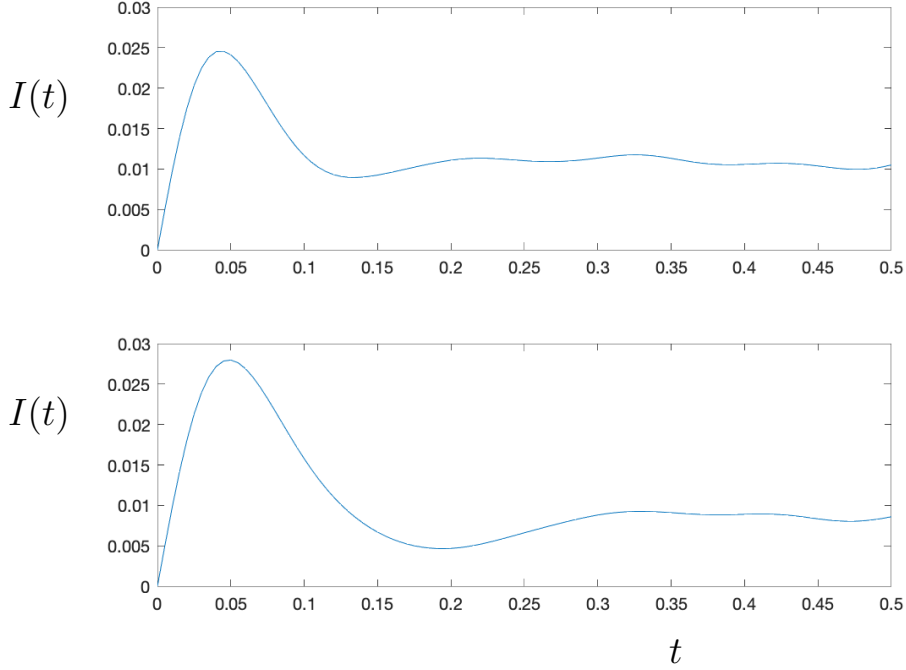
\begin{figure}
\centering
\begin{picture}(320,220)
\put(-10,10){\includegraphics[width=0.9\textwidth]{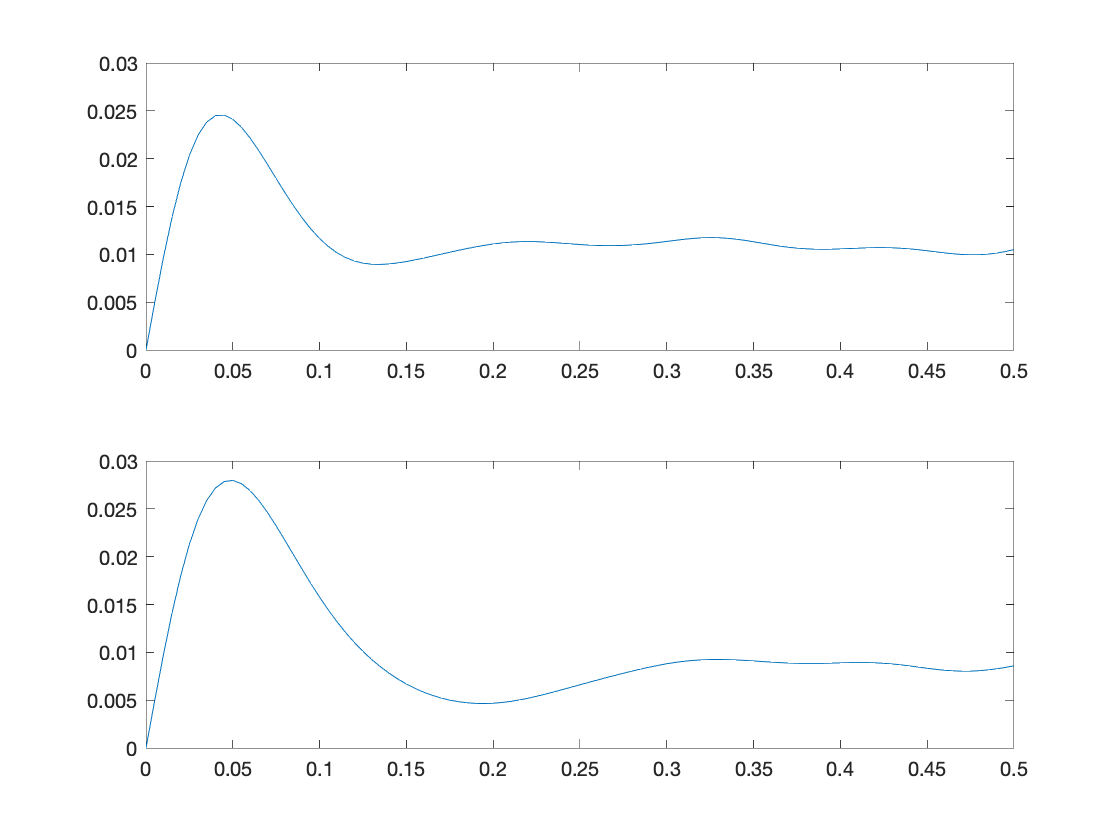}}
\put(-20,220){{\Large $I(t)$}}
\put(-20,100){{\Large $I(t)$}}
\put(230,10){{\Large $t $}}
\end{picture}
\caption{Simulation results for $I(t)=\int_0^t \psi(t') dt' $ as a function of non-dimensional time (see section \ref{MD}) for different conditions. The top panel shows the result for $n=0.64$, $T=1.2$, $C_{wf}=1$ and $\varepsilon_{wf}=0.4$; the bottom panel shows the result for $n=0.8$, $T=1.0$, $C_{wf}=0.6$ and $\varepsilon_{wf}=1$. }
\label{example_psi}
\end{figure}

\section{Comparison with Molecular Dynamics simulations}
\label{MD}
In this section we validate the model proposed here with MD simulation results.
\subsection{Methodology}
 Our MD simulations, performed using the LAMMPS software \cite{LAMMPS}, are also based on the GK formulation: they provide {\it equilibrium configurations} to calculate $R_K$ via (\ref{R_K_equation}) for a system comprising a dense liquid bounded by two fcc-structured walls in a slab geometry. 
In what follows, all quantities will be reported in  non-dimensional units using the characteristic time   $\tau_{LJ}=\sqrt{m_f{\sigma_{ff}}^2/\varepsilon_{ff}}$, the characteristic distance  $\sigma_{ff}$ and the  potential well depth $\varepsilon_{ff}$ associated with the fluid-fluid interaction. In all our simulations $C_{ww}=C_{ff}=1$, while $C_{wf}$ was varied in the range $0.4\leq C_{wf}\leq 1$ as will be discussed below.

\begin{figure}
\centering
\begin{picture}(320,250)
\put(-10,10){\includegraphics[width=0.85\textwidth]{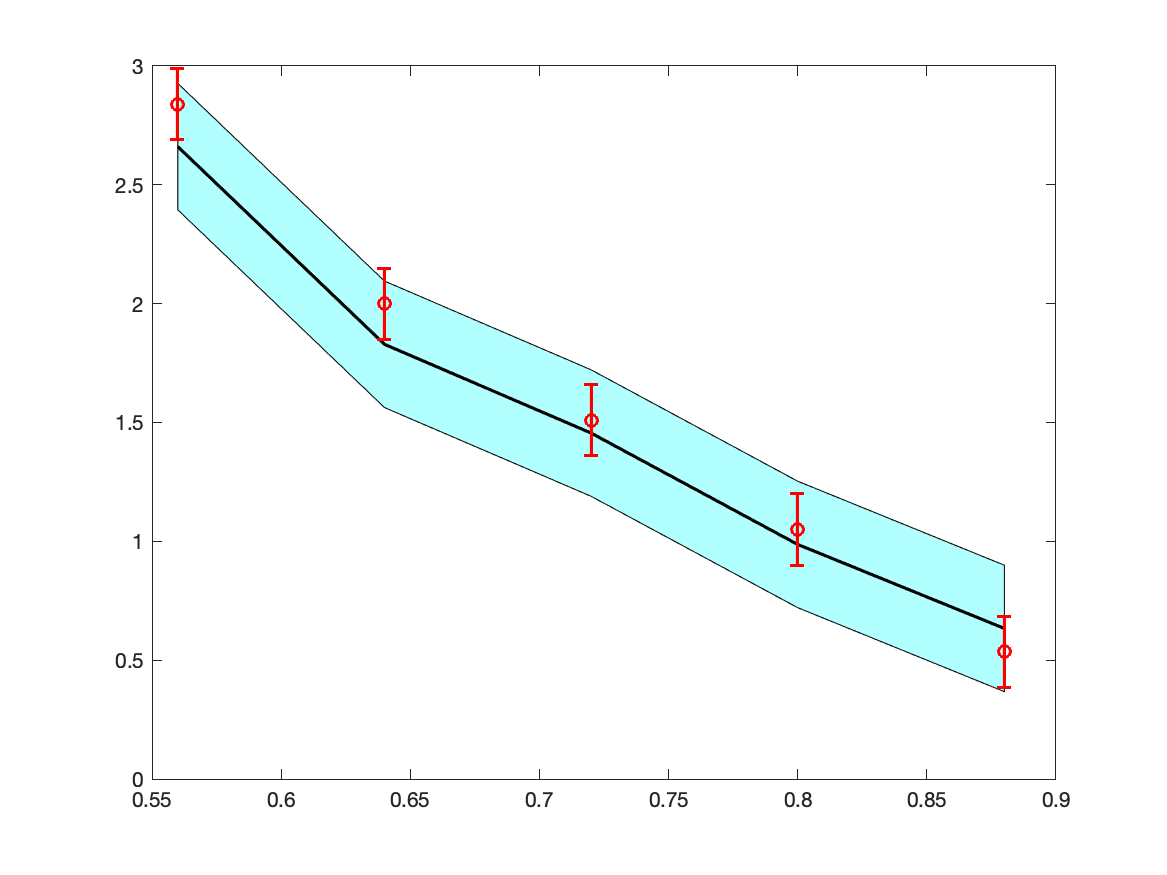}}
\put(0,180){{\Large $R_K$}}
\put(220,10){{\Large $n$}}
\end{picture}
\caption{Kapitza resistance as a function of fluid density at $T=1.5$, $\varepsilon_{wf}=1$ and $C_{wf}=0.6$. MD simulation results are shown in red symbols, while the predictions and uncertainty associated with model (\ref{final}) are shown by the black line and blue shading, respectively.}
\label{Rvsrho}
\end{figure}

In its nominal state, the fluid-solid system measured 30.8 LJ units in each of the two dimensions parallel to the wall, while the thickness of the fluid slab was $L=46.2$ units, in accordance with the recommendation of Alosious et al. \cite{Alosious} who found size effects to be negligible for fluid thicknesses greater than approximately 40 units; increasing the fluid slab thickness to 92.4 units did not produce any significant change in our results. The system size remained constant at the above  nominal values except when investigating the effect of the wall density (see below); during the latter simulations, system dimensions were variable with a maximum reduction of 30\%. 

Each wall consisted of a 7.71 unit thick FCC slab of atoms divided into three regions, each under different dynamics. The outermost region contained three atomic layers frozen in place. The middle region contained 7 atomic layers thermostated to the desired system temperature ($T$) via a Nos\'{e}–Hoover thermostat. The innermost region, in contact with the fluid, comprised of a single atomic layer under NVE dynamics. Unless otherwise stated, the wall density was fixed at $n_w$=1.09. In all simulations  $m_w=5$ ($m_f=1$), $\sigma_{wf}=1$ ($\sigma_{ff}=1$) and $\varepsilon_{ww}=4$  ($\varepsilon_{ff}=1$). A potential cutoff of 5 units was used.

After an equilibration period of 3000 LJ time units to ensure isothermal conditions, the resistance was calculated by numerical integration of the heat flux autocorrelation trace. Figure \ref{example_psi} shows the result of two such calculations at different conditions. The MD estimate for $R_K$ was calculated by inserting the result for $\int_0^\infty \langle q(t)q(0)\rangle dt$ into (\ref{R_K_equation}). 

Simulations were performed for a wide variety of conditions. Our results are presented in the next subsection and compared with the model (\ref{final}) with $D=12$; this value was chosen as the one which gives good overall agreement with the MD data.  Values for $M_\infty$ and $\langle \sum_i|{\bf F_w}|^2_i\rangle$ for use in (\ref{final}) were calculated from MD simulations, while the thermal conductivity of the fluid was calculated using the correlation of Galliero et al. \cite{Galliero}.
\begin{figure}
\centering
\begin{picture}(320,250)
\put(-10,10){\includegraphics[width=0.85\textwidth]{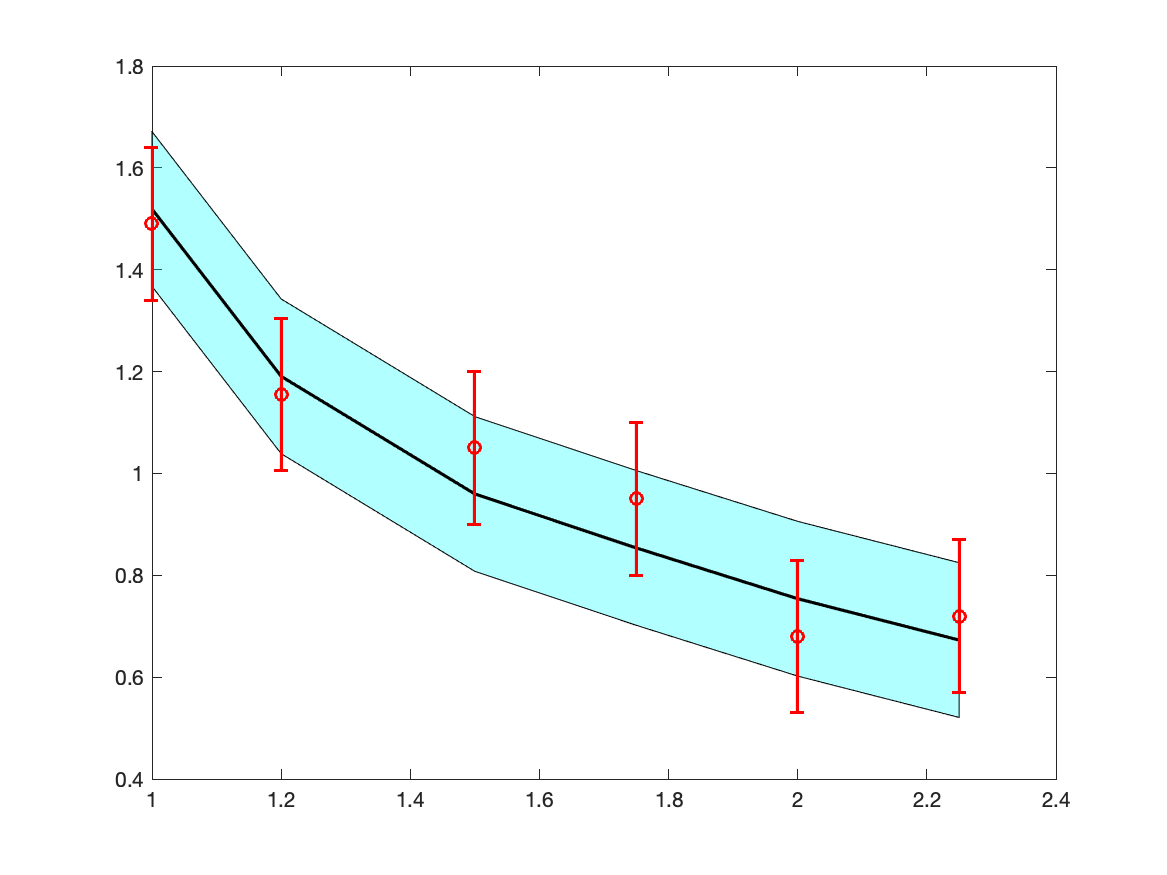}}
\put(0,180){{\Large $R_K$}}
\put(220,10){{\Large $T$}}
\end{picture}
\caption{Kapitza resistance as a function of temperature with $n=0.8$, $\varepsilon_{wf}=1$ and $C_{wf}=0.6$. MD simulation results are shown in red symbols, while the predictions and uncertainty associated with model (\ref{final}) are shown by the black line and blue shading, respectively.}
\label{RvsT}
\end{figure}
\subsection{Discussion of results}
Figure \ref{Rvsrho} shows the Kapitza resistance as a function of the fluid density, $n$, at $T=1.5$ with $\varepsilon_{wf}=1$ and $C_{wf}=0.6$. Here and in the above, $n$ denotes the nominal number density, defined as the ratio of the number of fluid atoms to the fluid volume, $V=A L$, where $L$ is the distance between the two walls. Although the density is not homogeneous in the wall vicinity \cite{Jerryinterface}, our results below suggest that use of the nominal density leads to a sufficiently accurate description. The monotonically decreasing resistance as a function of fluid density observed in the figure is in agreement with the very careful work of Alosious et al. \cite{Alosious}. This behavior can be understood by noting that the dependence of the relaxation timescale $\kappa/M_\infty$ on the fluid density is much weaker than that of the TSWF which is a monotonically increasing function of the fluid density. We also observe that the model (\ref{final}) matches the simulation results closely.

\begin{figure}
\centering
\begin{picture}(320,250)
\put(-10,10){\includegraphics[width=0.85\textwidth]{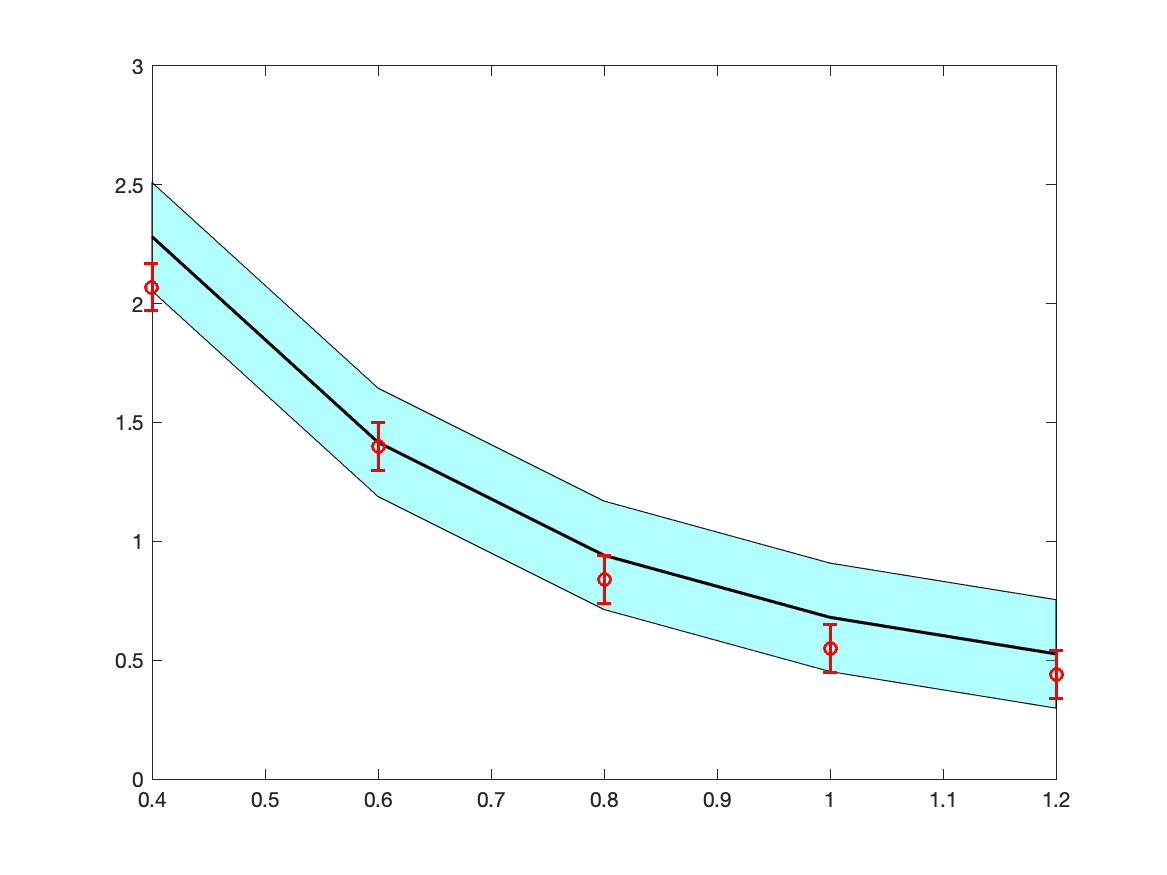}}
\put(0,180){{\Large $R_K$}}
\put(220,10){{\Large $\varepsilon_{wf}$}}
\end{picture}
\caption{Kapitza resistance as a function of the fluid-solid interaction strength at $T=1.2$, $n=0.64$ with  $C_{wf}=1$. MD simulation results are shown in red symbols, while the predictions and uncertainty associated with model (\ref{final}) are shown by the black line and blue shading, respectively.}
\label{Rvse}
\end{figure}

  The temperature dependence of the Kapitza resistance observed in figure \ref{RvsT} is in line with previous work \cite{song,Alosious} and in good agreement with model (\ref{final}) which can also provide more insight into the origins of this behavior.
  According to (\ref{final}), the temperature dependence of the Kapitza resistance is a result of competition between three factors, namely, explicit dependence of $R_K$ on $T$, temperature dependence of the relaxation timescale $\kappa/M_\infty$ and temperature dependence of the TSWF, the latter being a strongly increasing function of temperature. The monotonic decrease of the resistance as a function of temperature observed here and characterized in detail by Song and Min  \cite{song} can thus be understood by noting that, at these conditions, the relaxation timescale $\kappa/M_\infty$ is very weakly decreasing as a function of temperature.

\begin{figure}
\centering
\begin{picture}(320,250)
\put(-10,10){\includegraphics[width=0.85\textwidth]{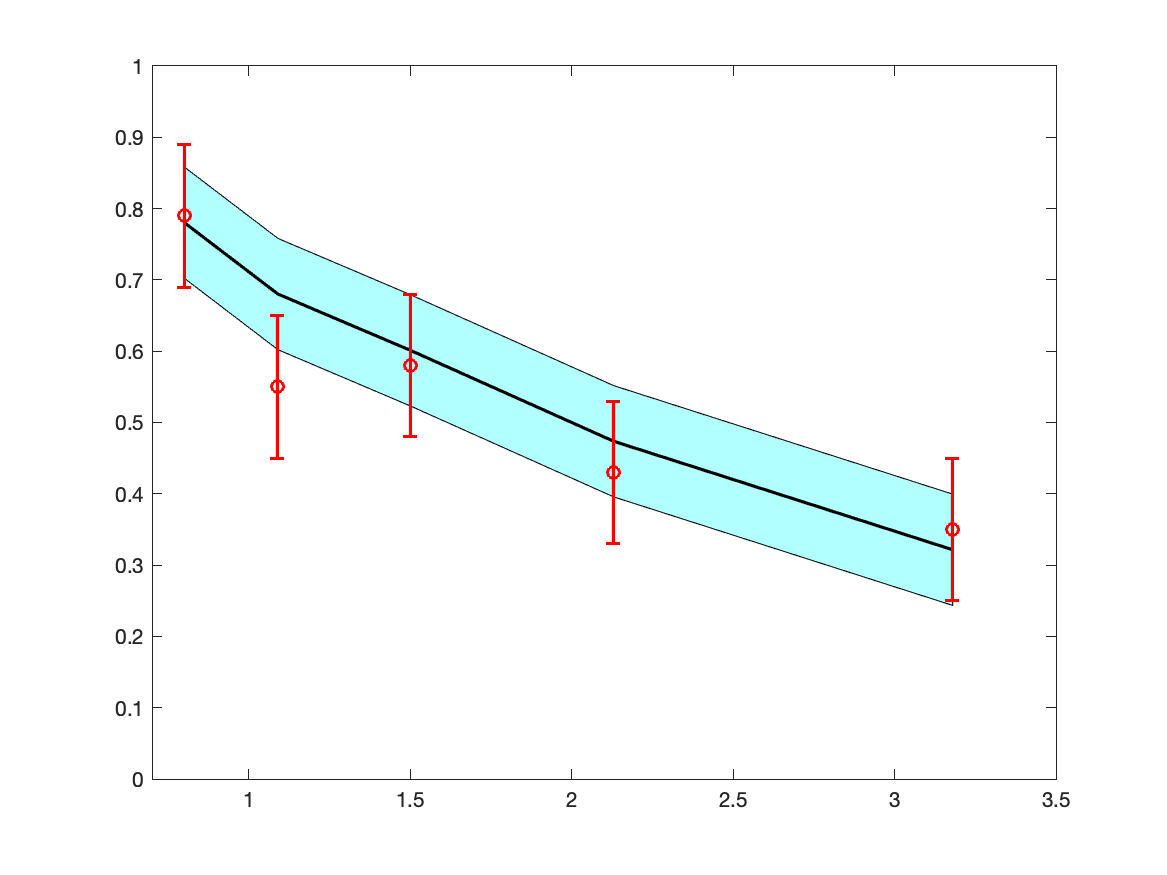}}
\put(0,180){{\Large $R_K$}}
\put(220,10){{\Large $n_w$}}
\end{picture}
\caption{Kapitza resistance as a function of the wall density at $T=1.2$, $n=0.64$ with $\varepsilon_{wf}=1$. MD simulation results are shown in red symbols, while the predictions and uncertainty associated with model (\ref{final}) are shown by the black line and blue shading, respectively.}
\label{Rvsc}
\end{figure}

The dependence of the Kapitza resistance on $\varepsilon_{wf}$ and $C_{wf}$ is simple to understand  qualitatively, since both these parameters only affect the TSWF. Specifically, the TSWF is a monotonically increasing function of  both $\varepsilon_{wf}$ and $C_{wf}$, making the Kapitza resistance a monotonically decreasing function of these parameters, as would be expected from previous work. Figures \ref{Rvse} and \ref{Rvsc} show this dependence for  $T=1.2$ and $n=0.64$. The agreement between the model and the simulation results is very good in both cases.

\section{Discussion}
\label{discussion}
We have developed and validated a {\it quantitative} model for the Kapitza resistance at a fluid-solid interface. The agreement between the model and simulation results is very encouraging and suggests that predictive calculations of the Kapitza resistance based on first-principles information on the fluid-solid system are possible, at least for simple fluids. Extension of this approach to more complex fluid-solid systems (see Ref. \cite{ohara} for example) is a clear next step.

In addition to an accurate quantitative description, the model offers physical insight into the primary factors that affect the behavior of the Kapitza resistance. The model shows that beyond the primary system properties ($m$, $T$), the resistance is determined by two composite properties, namely, the TSWF and the thermal relaxation time $I$. The TSWF, the primary factor affecting variations in the value of the resistance, is essentially a measure of the strength of interaction between the wall and fluid; a reduced interaction increases the thermal resistance and vice versa. This insight can be used to explain the recent observations \cite{Alosious2} on the dependence of $R_K$ on nanotube curvature and number of wall layers. The thermal relaxation time, or at least its approximation by $\kappa/D M_\infty$, is a property of the fluid only. Increasing the relaxation time reduces the thermal resistance and vice versa, although it should be noted that, as observed before, the ratio $\kappa/ M_\infty$ appears to be relatively insensitive to variations in the fluid state.

Although both the TWSF and $I$ can be calculated from MD simulations, as was the case in section \ref{MD}, it would be preferable if theoretical  expressions could be developed for their direct calculation. The approach of section \ref{TSWF} is one possible route; it can be completed by characterization of the liquid density profile in the wall vicinity and including the effect of wall structure. Our MD results show that expression (\ref{force-approx}) predicts the dependence of the TSWF on system parameters correctly, at least qualitatively. For the thermal modulus, we note that expressions for its mechanical analogues, the high-frequency shear and bulk moduli, $G_\infty$ and $K_\infty$ respectively, as a function of the fluid state, have been developed a long time ago by Zwanzig and Mountain \cite{Mountain}. Unfortunately, the only expression for the thermal modulus that the authors are aware of \cite{egelstaff} is very approximate. We finally note that since the thermal modulus is essentially a measure of the equilibrium fluctuations of the volume-averaged heat flux vector, it could also be useful for developing statistical error estimates for the heat flux vector in MD simulations using the procedure developed by Hadjiconstantinou et al. \cite{jcperror}. 

The most remarkable, perhaps, finding in this work is the relative success of the approximation  $I=\kappa/D M_\infty$ which enables the closed form expression (\ref{final}). Although physically motivated, further work is needed to understand the extent of its generality and develop robust approaches for calculating the value of coefficient $D$ from first principles.

The authors would like to thank Dr. Gerald Wang for many useful discussions. This material is based upon work supported by the Department of Energy, National Nuclear Security Administration under Award Number DE-NA0003965.

\printbibliography

@book{Sone,
  title={Molecular Gas Dynamics: Theory, Techniques, and applications},
  author={Sone, Y.},
  year={2007},
  publisher={Birkhauser}
}

@ARTICLE{Jerryslip,
   author = {G. J. Wang and N.G. Hadjiconstantinou},
   title = {A universal molecular-kinetic scaling relation for slip of a simple fluid at a solid boundary},
   journal = {Phys. Rev. Fluids},
    volume={4},
  pages={064291},
  year={2019},
}

@ARTICLE{Jerryinterface,
   author = {G. J. Wang and N.G. Hadjiconstantinou},
   title = {Molecular mechanics and structure of the fluid-solid interface in simple fluids},
   journal = {Phys. Rev. Fluids},
    volume={2},
  pages={094201},
  year={2017},
}

@ARTICLE{NGHpof,
   author = {N.G. Hadjiconstantinou},
   title = {The limits of {Navier-Stokes} theory and kinetic extensions for describing small-scale gaseous hydrodynamics},
   journal = {Phys. Fluids},
    volume={18},
  pages={111301},
  year={2006},
}

@ARTICLE{Alosious,
   author = {S. Alosious and S.K. Kannam and S.P. Sathian and B.D. Todd},
   title = {Prediction of {Kapitza} resistance at fluid-solid interfaces},
   journal = {J. Chem. Phys.},
  volume={151},
  pages={194502},
  year={2019},
}

@ARTICLE{Alosious2,
   author = {S. Alosious and S.K. Kannam and S.P. Sathian and B.D. Todd},
   title = {Nanoconfinement effects on the {Kapitza} resistance at water-{CNT} interfaces},
   journal = {Langmuir},
  volume={37},
  pages={2355-2361},
  year={2021},
}

@ARTICLE{Heyes,
   author = {D. M. Heyes},
   title = {Transport coefficients of {Lennard-Jones} fluids: A molecular dynamics and effective-hard-sphere treatment},
   journal = {Phys. Rev. B},
  volume={37},
  pages={5677--5696},
  year={1988},
}

@ARTICLE{Heyes2019,
   author = {D. M. Heyes and D. Dini and L. Castigliola and J. C. Dyre},
   title = {Transport coefficients of the {Lennard-Jones} fluid close to the freezing line},
   journal = {J. Chem. Phys.},
  volume={151},
  pages={204502},
  year={2019},
}

@ARTICLE{Peraud,
   author = {J-P. M. Peraud and N. G. Hadjiconstantinou },
   title = {Extending the range of validity of {Fourier's} law into the kinetic transport regime via asymptotic solution of the phonon {Boltzmann} transport equation},
   journal = {Phys. Rev. B},
  volume={93},
  pages={045424},
  year={2016},
}

@book{Hansen,
  title={Theory of simple liquids},
  author={J-P. Hansen and I.R. McDonald},
  year={2013},
  publisher={Academic Press}
}

@ARTICLE{chicou,
   author = {J-L. Barrat and F. Chiaruttini},
   title = {Kapitza resistance at the liquid-solid interface},
   journal = {Mol. Phys.},
  volume={101},
  pages={1605--1610},
  year={2003},
}

@ARTICLE{Beskok,
   author = {B. H. Kim  and A. Beskok and T. Kagin},
   title = {Molecular Dynamics simulations of thermal resistance at the liquid-solid interface},
   journal = {J. Chem. Phys.},
  volume={129},
  pages={174701},
  year={2008},
}

@ARTICLE{Mountain,
   author = {R. Zwanzig and R.D. Mountain },
   title = {High‐Frequency Elastic Moduli of Simple Fluids},
   journal = {J. Chem. Phys.},
  volume={43},
  pages={4464-4471},
  year={1965},
}

@book{egelstaff,
  title={An introduction to the liquid state},
  author={P.A. Egelstaff},
  year={1994},
  publisher={Clarendon Press}
}

@ARTICLE{jcperror,
   author = {N. G. Hadjiconstantinou and A. L. Garcia and M. Z. Bazant and G. He},
   title = {Statistical error in particle simulations of hydrodynamic phenomena},
   journal = {J. Comp. Phys.},
  volume={187},
  pages={274-297},
  year={2003},
}

@ARTICLE{BocquetAnnu,
   author = {N. Kavokine and R.R. Netz and L. Bocquet},
   title = {Fluids at the Nanoscale: From Continuum to Subcontinuum Transport},
   journal = {Annu. Rev. Fluid Mech.},
  volume={53},
  pages={377-410},
  year={2021},
}

@ARTICLE{jfm2021,
   author = {N. G. Hadjiconstantinou},
   title = {An atomistic model for the {Navier} slip condition},
   journal = {J. Fluid Mech.},
  volume={912},
  pages={A26},
  year={2021},
}

@ARTICLE{Keblinski2011,
   author = {Y. Wang and P. Keblinski},
   title = {Role of wetting and nanoscale roughness on
thermal conductance at liquid-solid interface},
   journal = {Appl. Phys. Lett.},
  volume={99},
  pages={073112},
  year={2011},
}

@ARTICLE{Cahill,
   author = {D.G. Cahill and W.K. Ford and K.E. Goodson and G.P. Mahan and A. Majumdar and H.J. Maris and R. Merlin and S.R. Phillpot},
   title = {Nanoscale Thermal Transport},
   journal = {J. Appl. Phys.},
  volume={93},
  pages={793},
  year={2003},
}

@ARTICLE{pohl,
   author = {E.T. Swartz and R.O. Pohl},
   title = {Thermal Boundary resistance},
   journal = {Rev. Mod. Phys.},
  volume={61},
  pages={605},
  year={1989},
}

@ARTICLE{kimbo,
   author = {H. Hu and Y. Sun},
   title = {Effect of nanopatterns on {Kapitza} resistance
at a water-gold interface during boiling: A
molecular dynamics study},
   journal = {J. Appl. Phys.},
  volume={112},
  pages={053508},
  year={2012},
}

@ARTICLE{Galliero,
   author = {M. Bugel and G. Galliero},
   title = {Thermal conductivity of the {Lennard-Jones} fluid: An empirical correlation},
   journal = {Chem. Phys.},
  volume={352},
  pages={249--257},
  year={2008},
}

@ARTICLE{LAMMPS,
   author = {S. Plimpton},
   title = {Fast Parallel Algorithms for Short-Range Molecular Dynamics},
   journal = {J. Comp. Phys.},
  volume={117},
  pages={1--19},
  year={1995},
}

@ARTICLE{ase,
   author = {K. Gordiz and A. Henry},
   title = {Phonon Transport at Crystalline {Si/Ge} Interfaces: The Role of Interfacial Modes of Vibration},
   journal = {Sci. Rep.},
  volume={6},
  pages={23139},
  year={2016},
}

@ARTICLE{Blady,
   author = {B. Ramos-Alvarado and S. Kumar and G.P. Peterson},
   title = {Solid-liquid thermal transport and its relationship with wettability and the interfacial liquid structure},
   journal = {J. Phys. Chem. Lett.},
  volume={7},
  pages={3497-3501},
  year={2016},
}

@ARTICLE{ohara,
   author = {Y. Guo and D. Surblys and H. Matsubara and Y. Kawagoe and T. Ohara},
   title = {Molecular dynamics study on the effect of long-chain surfactant adsorption on interfacial heat transfer between a polymer liquid and silica surface},
   journal = {J. Chem. Phys. C},
  volume={124},
  pages={27558--27570},
  year={2020},
}

@ARTICLE{song,
   author = {G. Song and C. Min},
   title = {Temperature dependence of thermal resistance at solid/liquid interface},
   journal = {Mol. Phys.},
  volume={111},
  pages={903--908},
  year={2013},
}

\end{document}